,

# Possible Display of Phason Mode of Electromagnons in TbMnO$_3$

I.E. Chupis

*B.Verkin Institute for Low Temperature Physics and Engineering, National Academy of Sciences of Ukraine, pr. Lenina 47, Kharkov 61103, Ukraine*

The interaction of light in terahertz frequency region with electromagnons in a sinusoidal incommensurate magnetic state in TbMnO$_3$ is studied. A significant change in the frequency dependence of the dielectric constant $\varepsilon_{zz}$ near the temperature of the phase transition from sinusoidal magnetic ordering to spiral spin structure with a spontaneous electric polarization is predicted. Phason mode of this phase transition is the mode of electromagnon with ME coupling proportional to the wave number of modulation structure.

TbMnO$_3$ is a multiferroic manganite with a simultaneous ordering of magnetic and ferroelectric subsystems (ferroelectromagnet) [1]. The value of the magnetoelectric (ME) interaction between subsystems is proportional to the magnetic and the dielectric susceptibilities therefore an appreciable values of ME effects can be expected only in the ferroelectromagnets with close temperatures of magnetic and ferroelectric transitions [2]. For example, in nickel-iodine boracite where the temperatures of magnetic and ferroelectric (FE) transitions coincide the change of the dielectric constant of the order of 30 % in a magnetic field was observed [3]. This result has not attracted a due attention and a long time only weak ME effects were observed in the ferroelectromagnets with no close temperatures of electric and magnetic orderings.

Recently in TbMnO$_3$ a strong magneto-dielectric coupling has been discovered: the induction of electric polarization by magnetic phase transition and the significant changes of the electric polarization and the dielectric constant in a magnetic field of the order of a several tesla [1]. Neutron scattering experiments have revealed that the FE phase transition in TbMnO$_3$ coincides with the magnetic transition from the longitudinal incommensurate structure to the incommensurate spiral structure [4]. The discovery of gigantic ME effects in TbMnO$_3$ meant the possibility of magnetic control of FE polarization and effective exploring of ferroelectromagnets for spin-electronics materials. Further gigantic ME effects were found in other ferroelectromagnets with close temperatures of FE and magnetic transitions [5-7]. Last time many different properties of ferroelectromagnets have been investigated. Possible applications of these materials in optoelectronics revived an interest in the searching for ME excitations. Hybrid FE-magnetic excitations in ferroelectromagnets (termed seignettomagnons) were predicted before [8]. Recently the first observation of a new type of ME hybrid excitations in TbMnO$_3$ and GdMnO$_3$ in terahertz frequency region has been reported [9]. In incommensurate antiferromagnetic (AF) phase without FE ordering ME excitations (named electromagnons) were excited by an a.c. electric field $\vec{e}$ directed in (X,Y) plane. It was shown that electromagnons may exist in systems with inhomogeneous magnetic structure without spontaneous FE polarization. The appearance of electromagnons below the temperature $T_N$ of magnetic transition in collinear incommensurate AF state was accompanied by the considerable change in the index of refraction. However no significant changes were observed in TbMnO$_3$ when passing the magnetic and FE phase transition temperature $T_l$ from collinear to noncollinear spiral magnetic state with a spontaneous electric polarization.

In this Letter we show that a possible significant change in the index of refraction near $T_l$ may be observed in a.c. electric field oriented along the direction Z of a spontaneous polarization $\vec{P}_0$. The phason mode of the phase transition at $T_l$ is a mode of electromagnons where ME coupling is proportional to the wave number of modulation structure.

Orthorhombic manganite TbMnO$_3$ (space group Pbnm) below $T_N \approx 41K$ has an exchange sinusoidal Ay type of AF structure with the orientations of the Mn$^{3+}$ spins and the wave vector of modulation structure $k \cong 0,28 b^*$ along Y axis [10]. The transition from sinusoidal Ay structure into the noncollinear spiral structure (Ay,Az) takes place at $T_l \approx 28K$. Simultaneously a static electric polarization $P_0$ along Z axis appears at $T = T_l$ [4]. The influence of the Tb$^{3+}$ spins on ME properties near $T_l$ is not significant as the terbium subsystem is paramagnetic ( the ordering temperature of  the terbium ions $T' \sim 7K$). The components Ay,Az of AF vector are parameters of the ordering at the phase transition temperatures T$_N$ and T$_l$ . Near the phase transition the Lagrange method is correct. The Lagrange function L=T- F where T is a kinetic energy, F is a functional of the Ginzburg-Landau. In our case we suppose:

$$T = V^{-1} \int d\vec{r} \frac{1}{2} \{\mu \dot{\vec{A}}^2 + \lambda \dot{\vec{P}}^2\}$$

$$F = V^{-1} \int d\vec{r} \{\frac{a}{2}\vec{A}^2 + \frac{w}{2} A_z^2 + \frac{1}{4} u \vec{A}^4 + \frac{1}{2}\gamma[(\partial_y A_y)^2 + (\partial_y A_z)^2] + \quad\quad (1)$$

$$\frac{1}{2}\alpha[(\partial_y^2 A_y)^2 + (\partial_y^2 A_z)^2] + \frac{b}{2} P^2 - Pe_z + \nu P_z(A_z \partial_y A_y - A_y \partial_y A_z)\}$$

Here the constants $u, b, \alpha$ are positive, $\gamma$ is negative, exchange constant $a = \xi(T - T_0)$. Anisotropy constant w >0 because of the orientation AF vector along the Y axis in a collinear phase below $T_N = T_0 + \gamma^2/4\alpha\xi$. The last term in (1) is the ME interaction of electric polarization with modulated AF structure.

Equilibrium AF states in TbMnO$_3$ can be presented by harmonic rows. Restricting in the first harmonic we suppose $A_{y0} = A_1 \cos ky, A_{z0} = A_2 \sin ky$. Substituting these expressions into (1) and minimizing $F$ with respect to $A_1, A_2, P$ and modulation vector $k$ we find the following values in the modulated collinear (1) and noncollinear (2) phases:

1. $A_2 = 0, P_{z0} = 0, A_1^2 = -4L_1/3u,$
   $L_1 = a - a_c < 0, k^2 = -\gamma/2\alpha, a_c = \gamma^2/4\alpha$

   (2)

2. $A_1^2 = (L_2 - 3L_1 + \varepsilon_1)/2u, A_2^2 = (L_1 - 3L_2 + \varepsilon_2)/2u,$
   $L_2 = a - a_c + w, \varepsilon_{1,2} = 2k^2\nu^2(3L_{1,2} - 5L_{2,1})(ub)^{-1}$
   $P_{z0} = P_0 = k\nu A_1 A_2 b^{-1}; 3L_1 < L_2 < L_1/3$

In (2) we consider ME interaction is weak, $k^2\nu^2 \ll ub$.

The phase transition of the second kind from 1 into 2 state takes place at $T = T_l < T_N, T_l \cong T_N - 3w(2\xi)^{-1} + 6k^2v^2w(ub\xi)^{-1}$. The values $A_2$ and $P_{zo}$ are zero at $T = T_l$.

Fluctuations of the AF vector $\vec{a} = \vec{A} - \vec{A}_0$ and electric polarization $\vec{p} = \vec{P} - \vec{P}_0$ are described by the Lagrange equations. Further we shall consider the excitation of these fluctuations in the collinear phase 1 by an electric field $e_z$ of electromagnetic wave running in the Y direction, $e_z, h_x \propto \exp[i(qy - \omega t)]$. The results received below remain if change $x \leftrightarrow y$. According to the Maxwell equations electric field $e_z$ excites electric polarization $p_z$, $\partial_y^2 e_z = -c^{-2}\omega^2(e_z + 4\pi p_z)$. The Lagrange equations in linear approximation connect the $p_z$ only with $a_z$:

$$\lambda \ddot{p}_z + (b - e_z)p_z + va_z \partial_y A_{y0} - vA_{yo} \partial_y a_z = 0$$
$$\mu \ddot{a} + (a + w + uA_{y0}^2 - \gamma \partial_y^2 + \alpha \partial_y^4)a_z + 2vp_z \partial_y A_{y0} + vA_{y0} \partial_y p_z = 0 \quad (3)$$

The decisions of the equations (3) we may search in the form $a(y,t) = \exp(iqy - i\omega t)\sum_n a_n \exp(inky)$. The terms with coefficient $v$ in (3) induce the first harmonic ($n = \pm 1$) while the function $A_{y0}^2(y)$ induces the second one. As the wave vector of electromagnetic wave $q \ll k$ we may neglect of the second harmonic terms in (3). Suppose $e_z(p_z) = e_0(p_0)\exp(iqy)$, then $p_0 = e_0(n^2 - 1)/4\pi$, where $n = qc/\omega$ is the index of refraction. The values of $p_0$ and the coefficients $a_1, a_{-1}$ of the first harmonic of the AF excitation $a_z$ are coupled

$$Bp_0 - iD_k a_{-1} - iD_{-k} a_1 = 0$$
$$iD_{-k} p_0 - Ra_{-1} + L_+ a_1 = 0 \quad (4)$$
$$iD_k p_0 + L_- a_{-1} - Ra_1 = 0$$

where

$$B = \lambda \omega^2 - b + 4\pi(n^2 - 1)^{-1}, D_k = A_1 v(k - q/2), R = uA_1^2/4$$
$$L_\pm = \mu \omega^2 - a - w - \gamma(q \pm k)^2 - \alpha(q \pm k)^4 - uA_1^2/2 \quad (5)$$

As $q \ll k$ we put q=0 in the $D_k, L_\pm$. Then $D_{-k} = -D_k, L_+ = L_-, a_{-1} = -a_1$. From (4) and (2) we receive

$$n^2 = \frac{(\Omega_1^2 - \omega^2)(\Omega_2^2 - \omega^2)}{(\omega_1^2 - \omega^2)(\omega_2^2 - \omega^2)} \quad (6)$$

Here

$$\omega_{1,2}^2 = \frac{1}{2}[\omega_p^2 + \omega_0^2 \mp \sqrt{(\omega_p^2 - \omega_0^2)^2 + 8k^2v^2A_1^2(\lambda\mu)^{-1}}]$$
$$\mu\omega_0^2 = w + 2(a - a_c)/3, \omega_p^2 = b\lambda^{-1}, \Omega_p^2 = (b + 4\pi)\lambda^{-1} \quad (7)$$

The expressions $\Omega_{1,2}^2$ are the same as $\omega_{1,2}^2$ with the change $\omega_p$ on $\Omega_p$.

The frequency $\omega_0$ is the AF frequency decreasing when $T \to T_l$; $\omega_p$ is the electro-dipole frequency. The $\omega_1$ and $\omega_2$ are the modes of the electromagnons with q=0. The coupling of AF and electro-dipole excitation in such electromagnon is proportional to the wave number of modulation structure $k$. It is a new type of electromagnon which exists only in a modulated magnetic state.

One can see from (7) that the electromagnon mode $\omega_1 = 0$ at $T = T_l$. It means that $\omega_1$ is a phason mode for the phase transition from collinear 1 into noncollinear 2 incommensurate magnetic states in TbMnO$_3$. The polariton spectrum (6) in the case $\omega_0 < \omega_p$ is schematically shown in Fig.1. There are two frequency regions near $\omega_1$ and $\omega_2$ where electromagnons interact with light resonantly. In the upper mode near $\omega_2$ the excitations of electric polarization dominate; in the lower mode near $\omega_1$ the AF excitations prevail. As the ME interaction is weak the values $\omega_1$ and $\Omega_1$ are near $\omega_0$, $\omega_1 < \Omega_1 < \omega_0, \omega_2 \approx \omega_p, \Omega_2 \approx \Omega_p$. The resonance interaction of light with the phason mode of electromagnons takes place in a narrow frequency interval

$$\omega_0 - \omega_1 = \frac{k^2 v^2 A_1^2}{\omega_0 \lambda \mu (\omega_p^2 - \omega_0^2)} \tag{8}$$

Near $T_l$ when $\omega_{1,0} \to 0$ the width of the interval (8) increases. The phase transition from collinear to noncollinear magnetic spiral structure at $T_l$ leads to a significant increase of the index of refraction (6), i.e. the dielectric constant $\varepsilon_{zz}$ at $\omega = \omega_1$.

May wait the increase of maximum value of imaginary part of $\varepsilon_{zz}$ simultaneously with the decrease of the frequency of this maximum when $T \to T_l$.

Notice that if a modulated structure is absent ($k=0$) we have $\Omega_1 = \omega_1$ and the refraction index (6) has only one resonance frequency $\omega_2$.

Summarizing, we have predicted the possibility to observe in TbMnO$_3$ a considerable change in the dielectric constant $\varepsilon_{zz}$ in terahertz frequency region near the temperature of the orientation phase transition $T_l$ in an incommensurate magnetic structure. A phason mode of this transition is a mode of the electromagnons existing only in a modulated magnetic structure.

<s></s>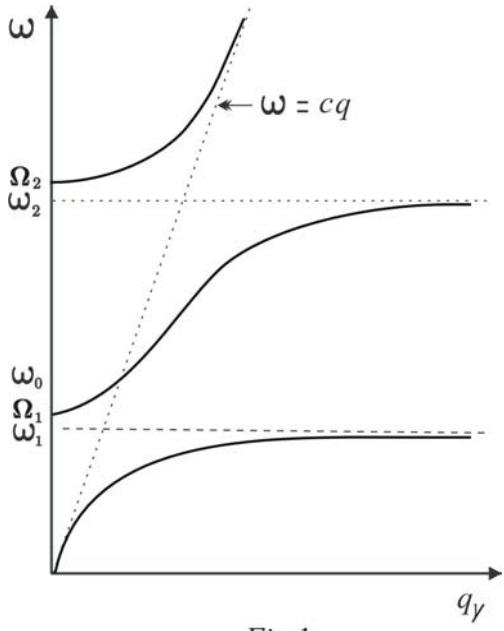

Fig.1

**Caption**

Fig.1 Sketch of polariton spectrum in a collinear incommensurate magnetic state (solid lines). The dashed line $\omega_1$ is the phason mode of electromagnon with dominated AF excitations; the dashed line $\omega_2$ corresponds to the electromagnon where electrodipole excitations prevail.